 \documentclass[aps,prb,twocolumn,groupedaddress,showpacs]{revtex4}

\usepackage{graphicx}

\begin{document}

\title{Structural and hyperfine characterization of $\sigma$-phase Fe-Mo alloys}

\author{J. Cieslak}
\email[Corresponding author: ]{cieslak@novell.ftj.agh.edu.pl}
\author{S. M. Dubiel}
\author{J. Przewoznik}
\author{J. Tobola}
\affiliation{AGH University of Science and Technology, Faculty of Physics and Applied Computer Science,
al. Mickiewicza 30, 30-059 Krakow, Poland}

\date{\today}

\begin{abstract}

A series of nine samples of $\sigma$-Fe$_{100-x}$Mo$_x$ with $44 \le x \le 57$ were
synthesized by a sintering method.  The samples were investigated experimentally
and theoretically.  Using X-ray diffraction
techniques structural parameters such as lattice constants, atomic positions
within the unit cell and populations of atoms over five different sublattices
were determined.  An information on charge-densities and electric field
gradients at particular lattice sites was obtained by application of the
Korringa-Kohn-Rostoker (KKR) method for electronic structure calculations.
Hyperfine quantities calculated with KKR
were successfully applied to analyze M\"ossbauer spectra measured at room
temperature.

\end{abstract}



\maketitle

\section{Introduction}

A $\sigma$-phase (space group P4$_2$/mnm, 30 atoms per unit cell), which belongs to
a class of the so-called Frank-Kasper phases \cite{Frank58} characterized by a high coordination
numbers (CN=12-15, for $\sigma$) can be formed in alloy systems by a solid state
reaction at elevated temperatures.  About 50 examples of $\sigma$ were up-to-now
found in such binary systems, including the Fe-Mo one.  A possibility of
occurrence of the $\sigma$-phase in Fe-Mo was first revealed in 1949 by
Goldschmidt \cite{Goldschmidt49}.  Studying a ternary system of Fe-Cr-Mo he had
identified a new phase stable in the temperature interval of 1180 - 1540 $^\circ$C.  The
phase had some similarities to the already-known $\sigma$-phase in Fe-Cr. Its
crystallographic structure and parameters were definitely identified five years
later \cite{Bergman54}.  Wilson was the first who determined the lattice parameters
and distribution of atoms over the five sublattices for $\sigma$-Fe$_{50}$Mo$_{50}$ sample
\cite{Wilson63}.  Further investigations depicting the structure and sites occupancy
by Fe and Mo atoms were continued by other investigators, too
\cite{Spooner63,Yakel83}.  Heijwegen determined borders for the existence of the
$\sigma$-phase in the Fe-Mo system \cite{Heijwegen71}.  Finaly, the phase diagram which is
regarded as the most actual was proposed by Guillermert \cite{Guillermet82}.

\begin{figure}[b]
\includegraphics[width=.49\textwidth]{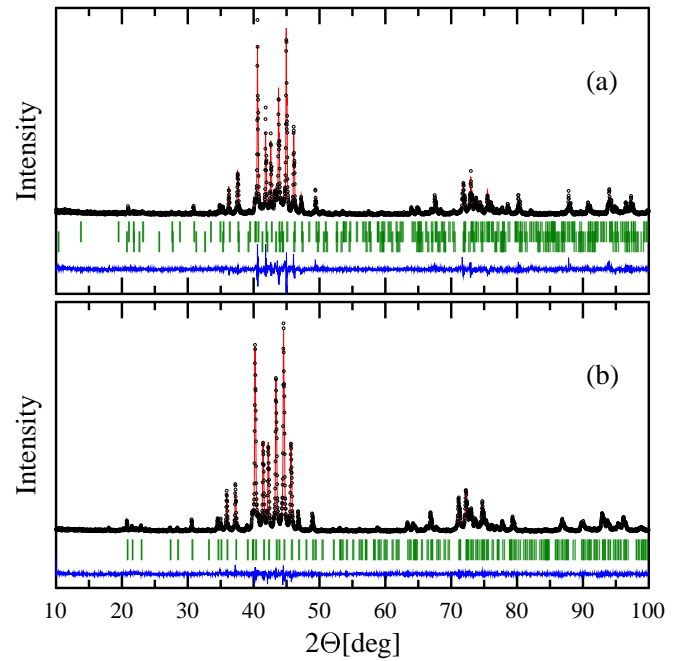}
\caption{(Online color)
Parts of of selected fitted X-ray diffractograms recorded at 294K on the $\sigma$-phase sample of
(a) Fe$_{55}$Mo$_{45}$ and (b) Fe$_{45}$Mo$_{55}$.
The solid line stays for the best-fit obtained with the procedure described in the text.
Peak positions for $\mu$ and $\sigma$ (a) as well as for pure $\sigma$-phase (b) are indicated.
A difference diffractograms are shown, too.
}
\label{fig1}
\end{figure}

In the Fe-Mo system, the $\sigma$-phase is one of three Frank-Kasper phases that can occur
in that system.
The other two ones are represented by $\mu$ (16 atoms per unit cell), and $R$ (54
atoms per unit cell).  They have different compositions and can be formed in
different ranges of temperature than $\sigma$ \cite{Arnfelt28,Sinha67}.  In the
available literature, there is little information relevant to physical
properties of the $\sigma$-phase in Fe-Mo.  Even the data on its crystallographic
structure in the whole concentration range of its occurrence is not complete.  In
these circumstances systematic studies aimed at filling this gap are justified, all
the more so such studies have been recently carried out for this phase in Fe-Cr
and Fe-V systems \cite{Dubiel11}.  A comparison of
corresponding data obtained for different binary alloy systems containing iron,
Fe-X, where the $\sigma$-phase can be formed, is of interest {\it per se}, and,
additionally, it may contribute to a better understanding of its formation
mechanism. The konowledge of the letter is of a great importance in the light of a detrimental effect
of a $\sigma$-phase presence in
stainless steels used in various branches of industry as valuable construction materials.
In this paper results of a systematic study of the lattice parameters, sites occupancies,
Fe-site charge-densities and electric field gradients obtained using various experimental
tools viz. X-ray diffraction (XRD) and M\"ossbauer spectroscopy (MS), as well as electronic structure
calculations (KKR method) are presented in the whole concentration range of the $\sigma$-phase existence
in the investigated alloys.

\begin{figure}[t]
\includegraphics[width=.49\textwidth]{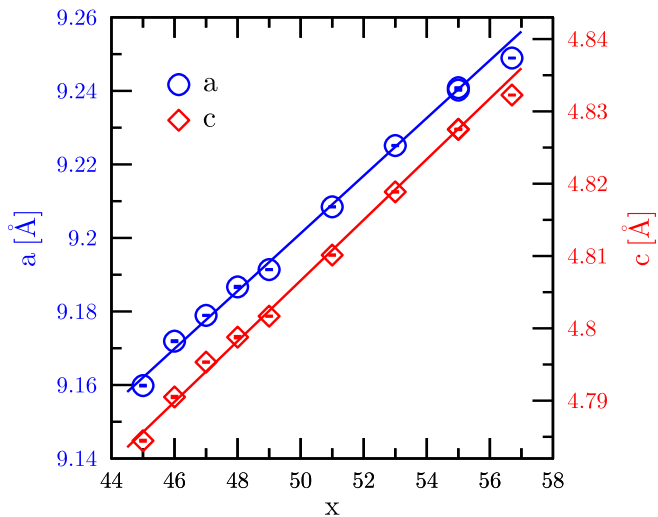}
\caption{(Online color)
Dependence of the lattice parameters $a$ (circles, blue) and c (diamonds, red) on
Mo-content, $x$, as determined from the X-ray diffractograms recorded at 294K.
}
\label{fig2}
\end{figure}

\section{Experimental}

It was not possible to obtain samples in a pure $\sigma$-FeMo alloys by using the
procedure that was previously successfully applied in the case of Fe-Cr and Fe-V alloys i.e.
an isothermal annealing of ingots obtained by melting constituents \cite{Dubiel11}.
The Fe-Mo ingots treated in that way were always two-phase, the second phase being $\mu$.
Successful was, however, a procedure described by Bergman \cite{Bergman54} viz.
a sintering.  For that purpose, powders of elemental Fe (99.9at\%) and Mo (99.95at\%)
were mixed together in adequate proportions.
Next, 2 g tablets for each composition were fabricated by pressing the
mixtures.  The tablets were subsequently isothermally
annealed at $1430^\circ$ C during 6
hours, and afterwords quenched into liquid nitrogen.  For XRD and M\"ossbauer
spectroscopic measurements, both carried out at room temperature (RT = 294K), the
pellets were attrited into powder in an agate mortar.  Altogether nine samples with the following
nominal concentration of Mo:  44, 45, 47, 48, 49, 51, 53, 55 and 57 at\% were
prepared.  Based on the XRD patterns, an example of which is shown in Fig.  1, it was found that all
samples except two with the lowest Mo content, and one with the highest
content of Mo were single-phase, namely $\sigma$.  The samples containing
nominally 44 and 45 at\% Mo had some admixture of $\mu$-phase (Fig. 1a), and the most
Mo-concentrated sample contained $\sim 4.2$wt\% of undissolved Mo-rich, {\it bcc} Fe-Mo phase.  In the latter
case the content of Mo in the $\sigma$-phase was estimated as equal to 56.7 at\%.

\begin{figure}[b]
\includegraphics[width=.49\textwidth]{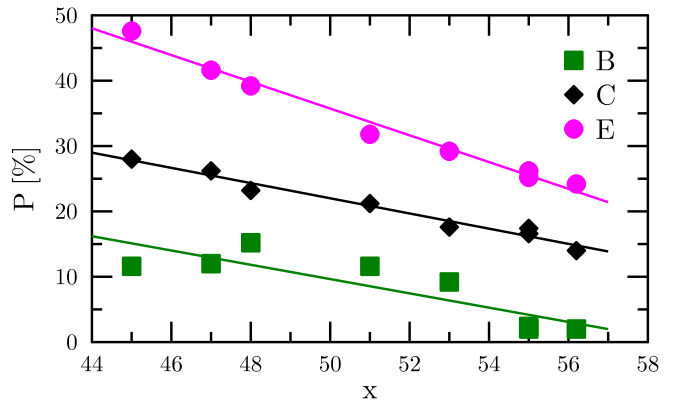}
\caption{(Online color)
Probability of finding Fe atoms at different lattice sites in the
$\sigma$-Fe$_{100-x}$Mo$_x$ compounds, $P$,  versus Mo concentration, $x$. Solid lines
stay for the linear fits to the data}
\label{fig3}
\end{figure}

\section{Results}

The powder XRD patterns were collected at RT with a D5000
Siemens diffractometer (using Cu K-$\alpha$ radiation and a graphite secondary
monochromator) from 10$^\circ$ to 140$^\circ$ in steps of 0.02$^\circ$ in 2$\theta$.  
Data were analyzed by the Rietveld method as implemented in the FULLPROF program
\cite{Rodriguez93} with 22 free
parameters:  6 of them were related to a background and positions of lines, 10
parameters represented sites occupancies and atomic positions in the unit cell.
Remaining 6 parameters described line widths, lattice constants and Debye-Waller
factors.  The analysis yielded values of the lattice constants, $a$ and $c$
(Fig.  2, Table 1), atomic positions (Table 2), as well as occupancies of the
sublattices (Fig.  3).  Concerning the lattice constants, as clearly
evidenced in Fig.  2, they exhibit a linear dependence on the Mo content, $x$.
Yet, the ratio $c/a$ = 0.52233(2) is rather $x$-independent.  For comparison,
Bergman found the value of 0.5237 for this ratio \cite{Bergman54}.  The values of the lattice
constants given in the literature \cite{Wilson63, Bergman54, Joubert08} are in line
with ours. This can be taken as indication that the nominal compositions of our samples can
be regarded as very close to the real ones.  Also the atomic positions found in
our analysis agree within error limits with those known from the literature
\cite{Bergman54, Joubert08}.  As far as the occupancy of the lattice sites is
concerned, the analysis of the XRD patterns gave evidence that sites A and D
were exclusively populated by Fe atoms, which is slightly different than in the
case of the $\sigma$-phase in Fe-Cr and Fe-V systems where the population of Fe
atoms at these sites was lying within 80-90\% range \cite{Cieslak08c}.  The remaining three sites were found
to be occupied by both Fe and Mo atoms, the former being in minority.  However, the
actual occupancy is concentration dependent and it decreases linearly with $x$ at
a rate characteristic of a given site. These results agree qualitatively with
those reported earlier\cite{Wilson63, Joubert08}.

\begin{figure}[t]
\includegraphics[width=.50\textwidth]{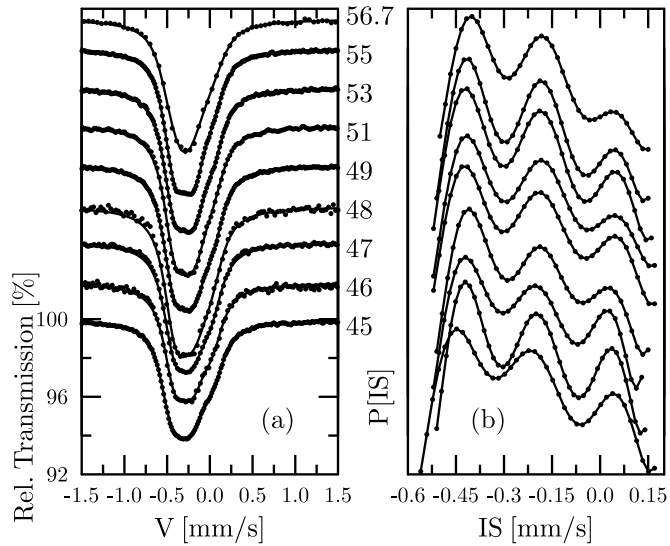}
\caption{
(a) $^{57}$Fe M\"ossbauer spectra recorded on a series of $\sigma$-Fe$_{100-x}$Mo$_x$ samples
at 294K and labeled with the corresponding $x$-values. The solid lines
are the best-fit to the experimental data. The derived isomer shift distribution curves,
shown in the same sequence as the spectra, are indicated in (b).
}
\label{fig4}
\end{figure}

$^{57}$Fe-site M\"ossbauer spectra, shown
in Fig. 4a, were recorded in a transmission geometry using a standard equipment and
Co/Rh source for the 14.4 keV gamma-radiation.  No traces of magnetism at RT were found
in all the spectra recorded at RT.  Despite it is known from the XRD experiments that Fe atoms are
present at all five sublattices, which means that each spectrum should be
composed of at least five subspectra. However, an unique analysis of the measured spectra in
terms of five components that could be associated with the five sublattices is not possible due to a
lack of a well-defined structure of them.
Instead, one can analyze them in terms of an isomer shift distribution, $ISD$, as explained elsewhere
\cite{Cieslak99}.  The $ISD$-curves obtained as a result of such analysis are presented
in Fig.  4b.  By integration of these curves, average values of the isomer shift, $\langle IS\rangle $,
were obtained (Table 1).  The dependence of the latter on the Mo content, $x$, is
illustrated in Fig.  5. It is clear than an increase of $x$ results in a linear
decrease of $\langle IS\rangle $ for all samples, except for those with the lowest $x$-values.  The
departure in the latter case is likely due to the fact that these two samples
had some small admixture of the $\mu$ phase. The decrease of $\langle IS\rangle $ (increase of the
Fe-site charge-density) with $x$ is rather weak, as it is equal to 0.0017 mm/s per
at\%.  For comparison, in the case of the $\sigma$-FeV alloys, the rate of $\langle IS\rangle $
decrease was almost two times higher \cite{Cieslak10a}.

\begin{table}[t] 
\caption{\label{table1} Lattice constants (in \AA), $a$ and $c$ as well as
the average isomer shift,$\langle IS \rangle$, (in $mm/s$, relative to the Co/Rh source), for all investigated
$\sigma$-Fe$_{100-x}$Mo$_x$ samples.
}
\begin{tabular}{|l|l|l|l|} \hline
$x$ & $a$             & $c$           & $\langle IS \rangle$          \\ \hline
45. & 9.1598(2)         & 4.7844(1)       & -0.242        \\ \hline
46. & 9.1719(2)         & 4.7905(1)       & -0.235        \\ \hline
47. & 9.1789(1)         & 4.7953(1)       & -0.234        \\ \hline
48. & 9.1866(2)         & 4.7987(1)       & -0.236        \\ \hline
49. & 9.1913(1)         & 4.8016(1)       & -0.236        \\ \hline
51. & 9.2084(1)         & 4.8101(1)       & -0.240        \\ \hline
53. & 9.2251(1)         & 4.8188(1)       & -0.242        \\ \hline
55. & 9.2402(1)         & 4.8275(1)       & -0.248        \\ \hline
56.7& 9.2489(1)         & 4.8322(1)       & -0.252        \\ \hline
\end{tabular}
\end{table}

\begin{table}[b] 
\caption{\label{table2} Atomic crystallographic positions, $x$, $y$, $z$, for the five
lattice sites of the Fe-Mo $\sigma$-phase. Since the values were weakly composition-dependent,
the average values over all samples are given.}
\begin{tabular}{|l|l|l|l|l|} \hline
Site& Wyckoff symbol    & x          & y             & z             \\ \hline
A   & 2i & 0          & 0             & 0             \\ \hline
B   & 4f & 0.3982(5)  & 0.3982(5)     & 0             \\ \hline
C   & 8i & 0.4635(4)  & 0.1297(5)     & 0             \\ \hline
D   & 8i & 0.7456(9)  & 0.0686(15)    & 0             \\ \hline
E   & 8j & 0.1819(3)  & 0.1819(3)     & 0.2469(11)    \\ \hline
\end{tabular}
\end{table}

A lack of a well-resolved structure of the spectra did not allow to decompose them uniquely
into the subspectra corresponding to the five sublattices, hence to determine
charge-densities and electric field gradients characteristic of these sublattices.
However, as demonstrated elsewhere \cite{Cieslak08b,Cieslak10a}, a knowledge of these
two spectral quantities can be obtained with the help of electronic structure calculations.
In the present case, the KKR calculations were carried out for 23 unit cells with different
configurations of Fe and Mo atoms on the sublattices A, B, C, D and E, keeping experimental
lattice constants in all cases. The configurations were chosen according to the following
two criteria:  (1) the probability of finding an Fe atom on a given site was as close as
possible to the one found experimentally for $\sigma$-Fe$_{56}$Mo$_{44}$, and (2) each possible
number of Fe atoms being the nearest-neighbors for a given lattice site, $NN_{Fe}$,
has been taken into account.

The charge and spin self-consistent Korringa-Kohn-Rostoker Green's function method \cite{mrs,cpa,stopa}
was here used to calculate the electronic structure of the Fe-Mo $\sigma$-phase. The crystal
potential of the muffin-tin form was constructed within the LDA framework, using the Barth-Hedin
formula for the exchange-correlation part. The experimental values of lattice constants and atomic
positions were used here. For fully converged crystal potentials electronic density of states (DOS),
total, site-composed and $l$-decomposed DOS (with $l_{max}=2$ for both types of atoms) were derived.
Fully converged results were obtained for $\sim 120$ special {\bf k}-points grid in the irreducible
part of Brillouin zone. DOS were computed using the tetrahedron {\bf k} space integration technique
and $\sim 700$ small tetrahedrons\cite{Kaprzyk86}. More details can be found elsewhere \cite{Cieslak08b,Cieslak10a},

The charge-densities, $\rho_{e}$, obtained as a result of these calculations were
analyzed as a function of the number of the nearest-neighbor Fe atoms, $NN_{Fe}$.
Graphical illustrations of the $\rho_{e}=f(NN_{Fe})$ dependences are presented in
Fig.  6. Two types of analysis were carried out.  First it was assumed the
$\rho_{e}$-$NN_{Fe}$ dependence, was linear, which is best seen for the sites D and E.  Taking into
account probabilities of each atomic configuration considered, the $\langle IS\rangle $-values were
calculated for each sublattice, and, finally, for each spectrum. The dependence of
$\langle IS\rangle $ on $x$ obtained within this approach is shown in Fig.  5 as a dashed line.
Obviously, it does not agree with the values derived from the $ISD$-curves
represented by circles in Fig.  5 which means the the linear dependence assumed in this approach was not correct.
In the second approach, the average isomer
shift for each subspectrum in a given spectrum was calculated as an arithmetic
average over all atomic configurations considered in the calculations. The average
$IS$ for the whole spectrum was computed as a weighted average over all five sublattices (the
weight of each subspectrum being its relative spectral area). The $\langle IS\rangle (x)$
dependence determined in this way is shown in Fig.  5 as a full line. It is clear that it
agrees very well with the experimentally found values.

\begin{figure}[t]
\includegraphics[width=.49\textwidth]{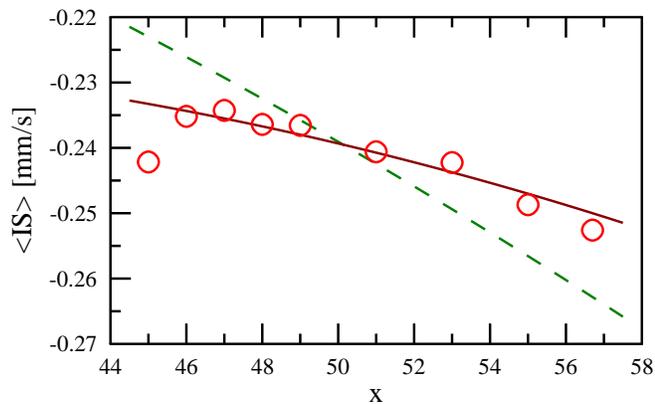}
\caption{(Online color)
The average isomer shift (relative to the Co/Rh source), $\langle IS \rangle$, versus Mo concentration, $x$, as measured (circles)
and as calculated (lines). The dashed line corresponds to the results obtained with approximation (1),
while the solid line with approach (2).
}
\label{fig5}
\end{figure}

\begin{figure}[b]
\includegraphics[width=.49\textwidth]{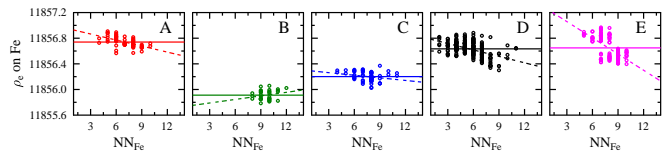}
\caption{(Online color)
Fe-site charge density, $\rho_e$, for five crystallographic sites versus $NN_{Fe}$.
Dashed lines correspond to the best linear fit to the data, whereas solid lines
represent average values of $\rho_e$}
\label{fig6}
\end{figure}

Based on the measured sublattice occupancies, calculated potentials and using
the extended point charge model as outlined elsehwere \cite{Cieslak08b}, it was also possible to determine
quadrupole splittings values, $QS$.  As shown in Fig.  7, the $QS$-values are
characteristic of a given lattice site, and for all of them but A, they exhibit
a weak linear increase with $x$ (for A there is a weak linear decrease).  The
smallest $QS$-value of 0.21 mm/s was found for the sublattice D, while the largest
one, 0.47 mm/s, for the site E.  The values of $QS$ for the other three sites have
intermediate values.  Noteworthy, similar values and relations were found for
the $\sigma$-phase in Fe-Cr and Fe-V systems \cite{Cieslak08b,Cieslak10a} which is reasonable in the light of the
same crystallographis structure.

The physical quantities determined in this study viz.  the site occupancies,
the charge-densities (isomer shifts) and the quadrupole splittings are sufficient to carry
out a proper analysis of the M\"ossbauer spectra in terms of the five subspectra corresponding to the five sublattices.
Only five free parameters were needed to successfully analyze the spectra.
Four of them i.e.  a background, spectral
area, line width (common for all five subspectra) and the isomer shift of one of
the subspectra depend on the conditions of the spectra measurements, so they
cannot be calculated.  The fifth free parameter was a proportionality constant
between the calculated electric field gradient component, $V_{zz}$, and the
corresponding spectral parameter viz. $QS$.  Examples of the spectra modeled in this
way together with the calculated subspectra corresponding to the particular lattice sites,
are illustrated in Fig.  8.  A very good quality of the fits achieved can be taken as a
proof that using the electronic structure calculations (KKR-method) the determined
spectral parameters adequately represent values of the
hyperfine parameters characteristic of the investigated samples. In other words, a combination of the M\"ossbauer-effect measurements with
the calculations of the electronic structure permited to propertly analyze the former in terms of the spectral parameters corresponding to
the different lattice sites.

\begin{figure}[t]
\includegraphics[width=.49\textwidth]{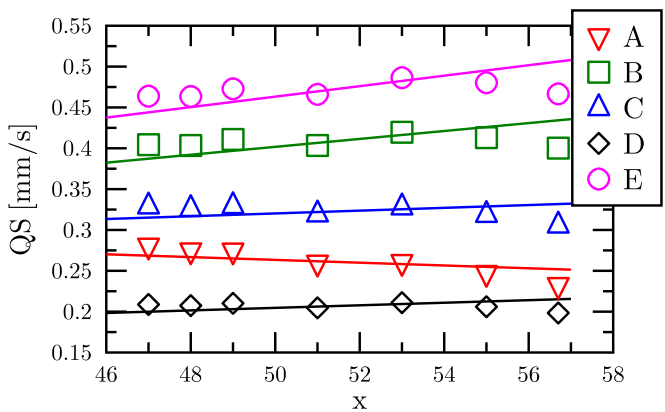}
\caption{(Online color)
Quadrupole splitting, $QS$, as determined
for each site and Mo concentration, $x$, from the analysis of
the measured spectra with the protocol described in Ref. \onlinecite{Cieslak08b}.
}
\label{fig7}
\end{figure}

\section{Conclusions}

The following conclusions can be drawn based on the results reported in this
paper:

1.  The pure $\sigma$-phase in the Fe-Mo alloy system can be obtained between
about 46 and 56.7 at\% Mo.

2.  Its lattice parameters $a$ and $c$ linearly increase with the Mo content, $x$.

3.  The $c/a$ ratio is concentration independent and equal to 0.52233(2).

4.  The sites A and D are 100\% occupied by Fe atoms, while the sites B, C and E
are populated by both types of atoms.The population of Fe atoms on the latter
three sites is in minority and it linearly decreases with $x$.

5.  The average isomer shift weakly decreases with $x$.

6.  The calculated charge-densities are characteristic of a given sublattice;
the highest one being at site D, and the lowest one at site B.

7.  The calculated quadrupole splittings are characteristic of a given
sublattice:  the largest one being at site E, and the smallest one at site D.

\begin{acknowledgments}
The results reported in this study were obtained within the project supported by the
  Ministry of Science and Higher Education, Warsaw (grant No. N N202 228837).
\end{acknowledgments}

\begin{figure}[b]
\includegraphics[width=.49\textwidth]{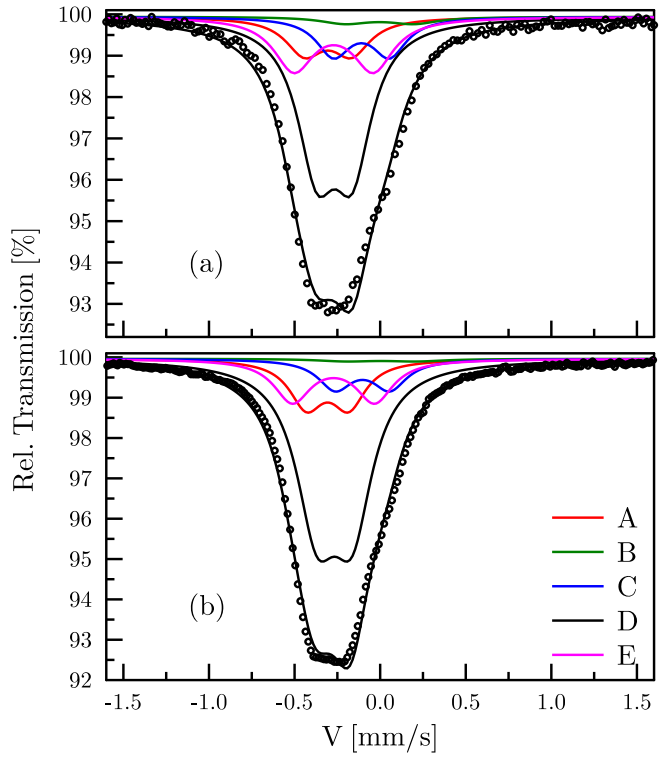}
\caption{(Online color)
$^{57}$Fe-site M\"ossbauer spectra recorded at
294K on two studied samples viz. with (a) $x=48$ and (b) $x=55$. The best-fit
spectrum and five subspectra are indicated by solid lines.}
\label{fig8}
\end{figure}

%


\begin{thebibliography}{58}


\expandafter\ifx\csname natexlab\endcsname\relax\def\natexlab#1{#1}\fi
\expandafter\ifx\csname bibnamefont\endcsname\relax
  \def\bibnamefont#1{#1}\fi
\expandafter\ifx\csname bibfnamefont\endcsname\relax
  \def\bibfnamefont#1{#1}\fi
\expandafter\ifx\csname citenamefont\endcsname\relax
  \def\citenamefont#1{#1}\fi
\expandafter\ifx\csname url\endcsname\relax
  \def\url#1{\texttt{#1}}\fi
\expandafter\ifx\csname urlprefix\endcsname\relax\def\urlprefix{URL }\fi
\providecommand{\bibinfo}[2]{#2}
\providecommand{\eprint}[2][]{\url{#2}}


\bibitem{Frank58}
  \bibinfo{author}{\bibfnamefont{C.}~\bibnamefont{Frank}}
  \bibnamefont{and}
  \bibinfo{author}{\bibfnamefont{J. S.}~\bibnamefont{Kasper}},
  \bibinfo{journal}{Acta Cryst.} \textbf{\bibinfo{volume}{11}},
  \bibinfo{pages}{184} (\bibinfo{year}{1958});
  \bibnamefont{ibid}
  \bibinfo{journal}{Acta Cryst.} \textbf{\bibinfo{volume}{12}},
  \bibinfo{pages}{4831} (\bibinfo{year}{1959}).


\bibitem{Goldschmidt49}
  \bibinfo{author}{{H.J. Goldschmidt}},
  \bibinfo{journal}{Research, Lond.}
   \textbf{\bibinfo{volume}{2}},
  \bibinfo{pages}{344}
 (\bibinfo{year}{1949}).

\bibitem{Bergman54}
  \bibinfo{author}{{G. Bergman}}
  and
  \bibinfo{author}{{D.P. Shoemaker}},
  \bibinfo{journal}{Acta Cryst.}
   \textbf{\bibinfo{volume}{7}},
  \bibinfo{pages}{857}
 (\bibinfo{year}{1954}).

\bibitem{Wilson63}
  \bibinfo{author}{{C.G. Wilson}}
  and
  \bibinfo{author}{{F.J. Spooner}},
  \bibinfo{journal}{Acta Cryst.}
   \textbf{\bibinfo{volume}{16}},
  \bibinfo{pages}{63}
 (\bibinfo{year}{1963}).

\bibitem{Spooner63}
  \bibinfo{author}{{F.J. Spooner}}
  and
  \bibinfo{author}{{C.G. Wilson}},
  \bibinfo{journal}{Acta Cryst.}
   \textbf{\bibinfo{volume}{17}},
  \bibinfo{pages}{1533}
 (\bibinfo{year}{1964}).

\bibitem{Yakel83}
  \bibinfo{author}{{H.L. Yakel}},
  \bibinfo{journal}{Acta Cryst.}
   \textbf{\bibinfo{volume}{B39}},
  \bibinfo{pages}{28}
 (\bibinfo{year}{1983}).

\bibitem{Heijwegen71}
  \bibinfo{author}{{C.P. Heijwegen}}
  and
  \bibinfo{author}{{G.D. Rieck}},
  \bibinfo{journal}{J. Less Common Metals}
   \textbf{\bibinfo{volume}{37}},
  \bibinfo{pages}{115}
 (\bibinfo{year}{1974}).

\bibitem{Guillermet82}
  \bibinfo{author}{{A.F. Guillermet}},
  \bibinfo{journal}{Bull. of Alloy Phase Diag.}
   \textbf{\bibinfo{volume}{3}},
  \bibinfo{pages}{359}
 (\bibinfo{year}{1982}).

\bibitem{Arnfelt28}
  \bibinfo{author}{{H. Arnfelt}},
  \bibinfo{journal}{Carnegie Scholarship Memoirs, Iron and Steel Inst.}
   \textbf{\bibinfo{volume}{17}},
  \bibinfo{pages}{6}
 (\bibinfo{year}{1928}).

\bibitem{Sinha67}
  \bibinfo{author}{{A.K. Sinha}},
  \bibinfo{author}{{R.A. Buckley}}
  and
  \bibinfo{author}{{W. Hume-Rothery}},
  \bibinfo{journal}{J. Iron and Steel Inst.}
   \textbf{\bibinfo{volume}{205}},
  \bibinfo{pages}{191}
 (\bibinfo{year}{1967}).

\bibitem{Dubiel11}
  \bibinfo{author}{\bibfnamefont{S.M. Dubiel}}
  \bibnamefont{and}
  \bibinfo{author}{\bibfnamefont{J. Cieslak}},
  \bibinfo{journal}{Crit. Rev. Sol. Stat. Mater. Sci.} \textbf{\bibinfo{volume}{36}},
  \bibinfo{pages}{191} (\bibinfo{year}{2011}).

\bibitem{Rodriguez93}
  \bibinfo{author}{\bibfnamefont{J. Rodriguez-Carjaval}},
  \bibinfo{journal}{Phisica B} \textbf{\bibinfo{volume}{192}},
  \bibinfo{pages}{55} (\bibinfo{year}{1993}).

\bibitem{Joubert08}
  \bibinfo{author}{\bibfnamefont{J.-M.}~\bibnamefont{Joubert}}
  \bibinfo{journal}{Progr. Mater. Sci.} \textbf{\bibinfo{volume}{53}},
  \bibinfo{pages}{528} (\bibinfo{year}{2008}).

\bibitem{Cieslak08c}
  \bibinfo{author}{\bibfnamefont{J.}~\bibnamefont{Cieslak}},
  \bibinfo{author}{\bibfnamefont{M.}~\bibnamefont{Reissner}},
  \bibinfo{author}{\bibfnamefont{S.~M.}~\bibnamefont{Dubiel}},
  \bibinfo{author}{\bibfnamefont{J.}~\bibnamefont{Wernisch}}
  \bibnamefont{and}
  \bibinfo{author}{\bibfnamefont{W.}~\bibnamefont{Steiner}},
  \bibinfo{journal}{J. Alloys Comp.} \textbf{\bibinfo{volume}{460}},
  \bibinfo{pages}{20} (\bibinfo{year}{2008}).

\bibitem{Cieslak99}
  \bibinfo{author}{{J. Cieslak}},
  \bibinfo{author}{{S.M. Dubiel}}
  and
  \bibinfo{author}{{B. Sepiol}},
  \bibinfo{journal}{Solid State Commun.}
   \textbf{\bibinfo{volume}{111}},
  \bibinfo{pages}{613}
 (\bibinfo{year}{1999})

\bibitem{Cieslak10a}
  \bibinfo{author}{\bibfnamefont{J.}~\bibnamefont{Cieslak}},
  \bibinfo{author}{\bibfnamefont{J.}~\bibnamefont{Tobola}},
  \bibnamefont{and}
  \bibinfo{author}{\bibfnamefont{S.~M.}~\bibnamefont{Dubiel}},
  \bibinfo{journal}{Phys.\ Rev.\ B} \textbf{\bibinfo{volume}{81}},
  \bibinfo{pages}{174203} (\bibinfo{year}{2010}).

\bibitem{Cieslak08b}
  \bibinfo{author}{{J. Cieslak}},
  \bibinfo{author}{{J. Tobola}},
  \bibinfo{author}{{S.~M. Dubiel}},
  \bibinfo{author}{{S. Kaprzyk}},
  \bibinfo{author}{{W. Steiner}}
  and
  \bibinfo{author}{{M. Reissner}},
  \bibinfo{journal}{J. Phys.: Condens. Matter.} \textbf{\bibinfo{volume}{20}},
  \bibinfo{pages}{235234} (\bibinfo{year}{2008}).

\bibitem[{\citenamefont{ed. by et~al.}(1992)\citenamefont{ed. by, Butler,
  Dederichs, Gonis, and Weaver}}]{mrs}
  \bibinfo{author}{\bibnamefont{ed. by}}, \bibinfo{author}{\bibfnamefont{W.~H.}
  \bibnamefont{Butler}},
  \bibinfo{author}{\bibfnamefont{P.}~\bibnamefont{Dederichs}},
  \bibinfo{author}{\bibfnamefont{A.}~\bibnamefont{Gonis}}, \bibnamefont{and}
  \bibinfo{author}{\bibfnamefont{R.}~\bibnamefont{Weaver}},
  \emph{\bibinfo{title}{Chapter III, in: Applications of Multiple Scattering
  Theory to Materials Science}}, vol. \bibinfo{volume}{253}
  (\bibinfo{publisher}{MRS Symposia Proceedings, MRS Pittsburgh.},
  \bibinfo{year}{1992}).


\bibitem[{\citenamefont{Stopa et~al.}(2004)\citenamefont{Stopa, Kaprzyk, and
  Tobola}}]{stopa}
  \bibinfo{author}{\bibfnamefont{T.}~\bibnamefont{Stopa}},
  \bibinfo{author}{\bibfnamefont{S.}~\bibnamefont{Kaprzyk}}, \bibnamefont{and}
  \bibinfo{author}{\bibfnamefont{J.}~\bibnamefont{Tobola}},
  \bibinfo{journal}{J. Phys.: Condens. Matter} \textbf{\bibinfo{volume}{16}},
  \bibinfo{pages}{4921} (\bibinfo{year}{2004}).


\bibitem[{\citenamefont{Bansil et~al.}(1999)\citenamefont{Bansil, Kaprzyk,
  Mijnarends, and Tobola}}]{cpa}
  \bibinfo{author}{\bibfnamefont{A.}~\bibnamefont{Bansil}},
  \bibinfo{author}{\bibfnamefont{S.}~\bibnamefont{Kaprzyk}},
  \bibinfo{author}{\bibfnamefont{P.~E.} \bibnamefont{Mijnarends}},
  \bibnamefont{and} \bibinfo{author}{\bibfnamefont{J.}~\bibnamefont{Tobola}},
  \bibinfo{journal}{Phys.\ Rev.\ B} \textbf{\bibinfo{volume}{60}},
  \bibinfo{pages}{13396} (\bibinfo{year}{1999}).

\bibitem{Kaprzyk86}
  \bibinfo{author}{\bibfnamefont{S.}~\bibnamefont{Kaprzyk}},
  \bibnamefont{and} \bibinfo{author}{\bibfnamefont{P.~E.}~\bibnamefont{Mijnarends}},
  \bibinfo{journal}{J.\ Phys.\ C} \textbf{\bibinfo{volume}{19}},
  \bibinfo{pages}{1283} (\bibinfo{year}{1986}).






\end{thebibliography}
\end{document}